\documentclass[preprint,eqsecnum]{revtex4}
\usepackage[centertags]{amsmath}
\usepackage{amssymb}
\usepackage{amsthm}

\begin{document}

\title{Noise-induced quantum transport}
\author{Pulak Kumar Ghosh, Debashis Barik and Deb Shankar Ray{\footnote{Email Address:
pcdsr@mahendra.iacs.res.in}}} \affiliation{Indian Association for
the Cultivation of Science, Jadavpur, Kolkata 700 032, India}

\begin{abstract}
We analyze the problem of directed quantum transport induced by
external exponentially correlated telegraphic noise. In addition
to quantum nature of the heat bath, nonlinearity of the periodic
system potential brings in quantum contribution. We observe that
quantization, in general, enhances classical current at low
temperature, while the differences become insignificant at higher
temperature. Interplay of quantum diffusion and quantum correction
to system potential is analyzed for various ranges of temperature,
correlation time and strength of external noise and asymmetry
parameters. A possible experimental realization of the observed
quantum effects in a superionic conductor placed in a random
asymmetric dichotomous electric field has been suggested.
\end{abstract}

\maketitle

\section{Introduction}

Extracting useful work from unbiased noise has been a topical
theme under 'ratchet effect' \cite{rei,jul,mag}. Because of its
extraordinary success in explaining experimental observations on
biochemical molecular motors active in muscle contractions
\cite{ser,val,how,tso}, observations on directed transport in
photoreflective and photovoltic materials \cite{gla,stu}, useful
application in the separation of particles \cite{rou}, theoretical
issues on symmetry consideration, Brownian motion and second law
of thermodynamics \cite{van,lef,sek,kam}, the problem has
attracted wide attention in recent years. Although the ratchet
effect can be achieved in a variety of ways the basic element of a
typical Brownian ratchet essentially concerns the breaking of the
detailed balance by an external periodic or fluctuating force
applied to the Brownian particle moving in a periodic potential.
This has been the subject a number of reviews and articles over
the last decade. We refer to [1-3] for details.

While in the recent context of ratchet a major emphasis is laid on
molecular pumps and motors in the realm of biophysics and
chemistry, it would seem that a Brownian particle being a
microscopic object \cite{van}, quantum effect is likely to make
its presence felt in appropriate situation, particularly at low
temperature. One thus expects the directed current or
rectification of noise to be important in transport of quantum
particles in quantum dots, wires related nanodevices
\cite{rei1,lin} and also in the context of superionic
conductors\cite{bru,gei,dic}. Furthermore such studies are
important also from the point of view of quantum-classical
correspondence. Based on the quantum Langevin equation our aim
here is to formulate a quantum ratchet problem where the external
time dependent modulation that drives the quantum system out of
equilibrium is a random telegraphic noise \cite{luc,doe,mie,kul}.
To incorporate the elements of quantum theory in a ratchet device
it is necessary to satisfy two basic requirements. First,
quantization of classical motion must not break the symmetry of
the ratchet device, or in other words, more specifically,
quantization should not bring in any additional tilt to the
potential or break its inversion symmetry or symmetry of the
detailed balance. Second, the forcing must be unbiased, so that
after appropriate averaging over ensemble or over the period of
space or time no directional component should remain. Any
approximation pertaining to the problem must conform to these
requirements in any correct quantum formalism.

 Keeping in view of the above considerations we
first formulate the quantum stochastic dynamics of an overdamped
particle and its approach to equilibrium. The quantum effects
appear due to the nonlinearity of the potential and quantum noise
of the heat bath. The introduction of the asymmetric telegraphic
noise breaks the symmetry of the detailed balance to produce
directed quantum transport. It is important to point out that the
quantum effects observed by Linke \textit{et al} \cite{lin} in
their experimental work on quantum dot device are due to tunneling
and wave reflection of electrons and this resulted in a reversal
of current as predicted by Reimann\cite{rei,rei1}. To explore the
quantum effects discussed in this paper we first note that the
quantum corrections have a different origin, e.g., the
nonlinearity of the potential. A possible candidate for the
experimental study of the present effect may be a superionic
conductor, e.g., AgI. This system has been traditionally used
\cite{bru,gei,dic} for measurement of current in presence of an
external electric field directly or in terms of frequency
dependent mobility. Typically in a superionic conductor like AgI,
$I^{-}$ ions form the lattice allowing the Ag$^{+}$ ions to move
in a periodic potential of the form $\cos\frac{2\pi x}{L}$, L
being the lattice spacing. The lattice vibrations contribute to
both the Langevin force as well as the frictional force on the
Ag$^{+}$ ions maintaining the detailed balance at the thermal
equilibrium. For slowly moving Ag$^{+}$ ions compared to lattice
vibrations, white noise approximation is sufficient. An
application of an external electric field at both ends of the
conductor which fluctuates randomly between two values in an
asymmetric way obeying the prescribed noise statistics, is
expected to result in an observable current. In what follows from
the present analysis is that apart from the low temperature
contribution due to deep tunneling, the nonlinearity of the
periodic pendulum potential $\cos\frac{2\pi x}{L}$ (contribution
beyond harmonic) contributes significantly to quantum corrections.
The typical experimental
 parameters for the measurement of current in a
superionic conductor has been given elsewhere\cite{bru}. It is
also expected that quantum dots where the confining potential is
truly periodic and nonlinear of similar type can offer themselves
as good candidates for these studies.

The outlay of the paper is as follows: we first derive in Sec.II
the basic equation describing quantum stochastic dynamics on a
general footing followed by an overdamped description.
Thermodynamic consistency has been stressed to avoid the pitfall
of fictitious current generation. In Sec.III we show how an
external dichotomous asymmetric noise can break the condition of
detailed balance inducing a directed transport. Two limiting cases
have been worked out in detail with a typically nonlinear
periodic(cosine) potential as an example. The paper is concluded
in Sce.IV.

\section{A quantum system in a spatially periodic potential at equilibrium}

\subsection{General aspects}
We consider a particle of mass $m$ moving in a periodic classical
potential $V(x)$. The particle is coupled to a set of harmonic
oscillators of unit mass acting as a bath. This is represented by
the following system-reservoir Hamiltonian \cite{db1,db2,db3}

\begin{equation}\label{2.1}
\hat{H}=\frac{\hat{p}^2}{2 m}+V(\hat{x})+\sum_{j=1}^N \left\{
\frac{\hat{p}^2_j}{2}+\frac{1}{2} \kappa_j (\hat{q}_j-\hat{x})^2
\right\}
\end{equation}

Here $\hat{x}$ and $\hat{p}$ are the coordinate and momentum
operators of the particle and $\{\hat{q}_j, \hat{p}_j\}$ are the
set of coordinate and momentum operators for the reservoir
oscillators coupled linearly through the coupling constants
${\kappa}_j (j=1,2,...)$. For the spatially periodic potential,
we have $V(x)=V(x+L)$,where $L$ is the length of the period.The
coordinate and momentum operators follow the usual commutation
rules $\{\hat{x}, \hat{p}\}=i\hbar$ and $\{\hat{q}_i,
\hat{p}_j\}=i\hbar{\delta}_{ij}$. Eliminating the bath degrees of
freedom in the usual way we obtain the operator Langevin equation
for the particle

\begin{equation}\label{2.2}
m \ddot{\hat{x}}+\int^{\overline t}_0
d\overline{t'}\gamma(\overline{t}-\overline{t'})\dot{\hat{x}}(\overline{t'})+V'(\hat{x})
= \hat{\Gamma}(\overline{t})
\end{equation}

(Overdots refers to differentiation with respect to time
$\overline{t}$) where noise operator $\hat{\Gamma}(\overline{t})$
and the memory kernel are given by

\begin{equation}\label{2.3}
\hat{\Gamma}(\overline{t}) =
\sum_j\left[\{\hat{q}_j(0)-\hat{x}(0)\}\kappa_j\cos\omega_j\overline{t}
+\kappa_j^{1/2}\hat{p}_j(0) \sin\omega_j\overline{t}\right]
\end{equation}

and

\begin{equation}\label{2.4}
\gamma(\overline{t}) =\sum_j\kappa_j\cos\omega_j\overline{t}
\end{equation}

respectively, with $\kappa_j=\omega_j^2$

Following Ref. [21-23] we then carry out a quantum mechanical
average $\langle...\rangle$ over the product separable bath modes
with coherent states and the system mode with an arbitrary state
at $\overline{t} =0$ in Eq.(\ref{2.2}) to obtain a generalized
quantum Langevin equation as

\begin{equation}\label{2.5}
m \ddot{\overline{x}} +
\int_0^{\overline{t}}d\overline{t'}\gamma(\overline{t}-\overline{t'})
\dot{\overline{x}}(\overline{t'})+\overline{V'}(\overline{x})=
\overline{\Gamma}(\overline{t}) + \overline{Q}(\overline{x}
,\overline{\langle\delta\hat{x}^n\rangle})
\end{equation}

where the quantum mechanical mean value of the position operator
$\langle\hat{x}\rangle =\overline{x}$ and

\begin{equation}\label{2.6}
\overline{Q}(\overline{x}
,\overline{\langle\delta\hat{x}^n\rangle)} =
\overline{V'}(\overline{x}) - \overline{\langle
V'(\hat{x})\rangle}
\end{equation}

which by expressing  $ \hat{x}(\overline{t}) =
\overline{x}(\overline{t}) + \delta\hat{x}(\overline{t})$ in
$V(\hat{x})$ and using a Taylor series expansion around
$\overline{x}$ may be rewritten as

\begin{equation}\label{2.7}
\overline{Q}(\overline{x}
,\overline{\langle\delta\hat{x}^n\rangle}) =-\sum_{n\geq
2}\frac{1}{n!}\overline{V}^{n+1}(\overline{x})\overline{\langle\delta\hat{x}^n\rangle}
\end{equation}
The above expansion implies that the nonzero anharmonic terms
beyond $n\geq 2$ contain quantum dispersions
$\overline{\langle\delta\hat{x}^n\rangle}$. Although we develop
this section in general terms, we are specifically concerned here
typically with periodic nonlinear potentials of the type
$\sin\frac{2\pi x}{L}$ or $\cos\frac{2\pi x}{L}$ or their linear
combinations and the like which have been used earlier in several
contexts. The nonlinearity of the potential is an important source
of quantum correction in addition to the quantum noise of the heat
bath. The calculation of $\overline{Q}$ rests on the quantum
correction terms $\overline{\langle{\delta\hat{x}^n}\rangle}$
which one determines by solving a set of quantum correction
equations as given in the next section. Furthermore the c-number
Langevin force is given by

\begin{equation}\label{2.8}
\overline{\Gamma}(\overline{t}) =
\sum_j\left[\langle\hat{q}_j(0)\rangle-\langle\hat{x}(0)\rangle
\kappa_j\cos\omega_j\overline{t} +\kappa_j^{1/2}\hat{p}_j(0)
\sin\omega_j\overline{t}\right]
\end{equation}

which must satisfy noise characteristics of the bath at
equilibrium ,

\begin{eqnarray}
\langle \overline{\Gamma}(\overline{t}) \rangle_S & = &
0\label{2.9}\\
\langle \overline{\Gamma}(\overline{t})
\overline{\Gamma}(\overline{t'}) \rangle_S &=& \frac{1}{2} \sum_j
\kappa_j\; \hbar \omega_j \left( \coth \frac{\hbar \omega_j}{2 k
T} \right) \cos \omega_j (\overline{t}-\overline{t'})\label{2.10}
\end{eqnarray}

Eq.(\ref{2.10}) expresses the quantum fluctuation-dissipation
relation. The above conditions(2.9-2.10) can be fulfilled provided
the initial shifted co-ordinates
$\{\langle\hat{q}_j(0)\rangle-\langle\hat{x}(0)\rangle\}$ and
momenta $\langle{\hat{p}_j}(0)\rangle$ of the bath oscillators are
distributed according to the canonical thermal Wigner distribution
\cite{wig,hil} of the form

\begin{equation}\label{2.11}
P_j([\langle\hat{q}_j(0)\rangle-\langle\hat{x}(0)\rangle],
\langle\hat{p}_j(0)\rangle) = N \exp\left\{-\;\frac{\frac
{1}{2}\langle\hat{p}_j(0)\rangle^2 + \frac
{1}{2}\kappa_j[\langle\hat{q}_j(0)\rangle-\langle\hat{x}(0)\rangle]^2}
{\hbar\omega_j[\overline{n}(\omega_j) + \frac{1}{2}]}\right\}
\end{equation}

so that the statistical averages $\langle...\rangle_s $ over the
quantum mechanical mean value $O$ of the bath variables are
defined as

\begin{equation}\label{2.12}
\langle O_j \rangle_s=\int O_j\; P_j\; d\langle
\hat{p}_j(0)\rangle\;
d\{\langle\hat{q}_j(0)\rangle-\langle\hat{x}(0)\rangle\}
\end{equation}

Here $\overline{n}(\omega)$ is given by Bose-Einstein
distributions $(e^{\frac{\hbar\omega}{kT}}-1)^{-1}$. $P_j$ is the
exact solution of Wigner equation for harmonic oscillator
\cite{wig,hil} and forms the basis for description of the quantum
noise characteristics of the bath kept in thermal equilibrium at
temperature $T$. In the continuum limit the
fluctuation-dissipation relation (\ref{2.10}) can be written as

\begin{equation}\label{2.13}
\langle\overline{\Gamma}(\overline{t})\overline{\Gamma}(\overline{t'})\rangle
= \frac{1}{2}\;\int_0^\infty d\omega\;
\kappa(\omega)\;\rho(\omega)\; \hbar\omega\;
\coth({\frac{\hbar\omega}{2kT}})\;\cos{\omega(\overline{t}-\overline{t'}})
\end{equation}

where we have introduced the density of the modes $\rho(\omega)$.
Since we are interested in the Markovian limit in the present
context, we assume  $\kappa(\omega)\rho(\omega)
=\frac{2}{\pi}\gamma$, Eq.(\ref{2.13}) then yields

\begin{equation}\label{2.14}
\langle\overline{\Gamma}(\overline{t})\overline{\Gamma}(\overline{t'})\rangle
=2 \overline{D}_q\delta(\overline{t}-\overline{t'})
\end{equation}

with

\begin{equation}\label{2.15}
\overline{D}_q
=\frac{1}{2}\gamma\hbar\omega_0\coth{\frac{\hbar\omega_0}{2kT}}
\end{equation}

 $\omega_0$ refers to static frequency limit.
 Furthermore from Eq.(\ref{2.4}) in the continuum limit we have

\begin{equation}\label{2.16}
\gamma(\overline{t}-\overline{t'}) =
\gamma\;\delta(\overline{t}-\overline{t'})
\end{equation}

$\gamma$ is the dissipation constant in the Markovian limit. In
this limit Eq.(\ref{2.5}) therefore reduces to

\begin{equation}\label{2.17}
m \ddot{\overline{x}} + \gamma
\dot{\overline{x}}+\overline{V'}(\overline{x})=
\overline{\Gamma}(\overline{t }) + \overline{Q}(\overline{x}
,\overline{\langle\delta\hat{x}^n\rangle})
\end{equation}

It is useful to work with dimensionless variables for the present
problem to keep track of the relations between the scales of
energy, length and time. The period $L$ of the periodic potential
$V(x)$ determines in a natural way the characteristic length scale
of the system. Therefore the position of the Brownian particle is
scaled as

\begin{equation}\nonumber
x=\overline{x}/L
\end{equation}

Next we consider the timescales of the system. In absence of the
potential and the noise term the velocity of the particle
$\dot{x}(\overline{t})\sim \exp(-\overline{t}/\tau_L)$ with
$\tau_L=m/\gamma$, which represents the correlation time scale of
the velocity the Brownian particle. To identify the next
characteristic time $\tau_0$ we consider the deterministic
overdamped motion due to the potential as
$\gamma\frac{d\overline{x}}{d\overline{t}} =
-\frac{d\overline{V}(\overline{x})}{d\overline{x}}$. Then $\tau_0$
is determined from
$\gamma\frac{L}{\tau_0}=-\frac{\overline{\Delta V}\ }{L}$ as
$\tau_0=\frac{\gamma L^2}{\overline{\Delta V}}$ where
$\overline{\Delta V} $ is the barrier height of the original
potential. Hence time is scaled as $t =
\frac{\overline{t}}{\tau_0}$.  Furthermore the potential, the
noise and the quantum correction terms are re-scaled as
$V(x)=\overline{V}(\overline{x})/\Delta\overline{V}$, $
\Gamma(t)=\overline{\Gamma}(t)/(\Delta\overline{V}/L)$ and
$\overline{Q}/(\Delta\overline{V}/L)$, respectively.

Hence dimensionless quantum Langevin equation reads as

\begin{equation}\label{2.18}
\mu^*\ddot{x}+\dot{x}=f(x)+\Gamma(t)
\end{equation}

Here over-dot(.) refers to differentiation with respect to scaled
time $t$. Dimensionless mass
$\mu^*=\frac{m}{\gamma\tau_0}=\frac{\tau_L}{\tau_0}$ and

\begin{equation}\label{2.19}
f(x)=-V'(x)+ Q(x ,\langle\delta\hat{x}^n\rangle)
\end{equation}

The noise properties of the quantum bath are then rewritten as

\begin{equation}\nonumber
\langle\Gamma(t)\rangle_s = 0
\end{equation}
\begin{equation}\nonumber
\langle\Gamma(t)\Gamma(t')\rangle_s =2 D_q\delta(t-t')
\end{equation}
where

\begin{equation}\nonumber
D_q=\frac{\frac{1}{2}\hbar\omega_0\coth{\frac{\hbar\omega_0}{2kT}}}{\overline{\Delta
V }}
\end{equation}

 The Fokker-Planck equation corresponding to Eq.(2.18) is given
by

\begin{equation}\label{2.20}
\frac{\partial P(x,\dot{x},t)}{\partial
t}=\left[-\frac{\partial}{\partial
x}\dot{x}+\frac{\partial}{\partial
\dot{x}}\left(\frac{\dot{x}}{\mu^*}-\frac{f(x)}{\mu^*}\right)
+\frac{D_q}{{\mu^*}^2}\frac{\partial^2}{\partial
x^2 }\right]P(x,\dot{x},t)
\end{equation}

The above equation can be solved in the stationary state. The
stationary probability density is

\begin{equation}\label{2.21}
P_{st}(x,\dot{x})=N \exp\left[-\frac{\mu^*\dot{x}^2}{2}+ \int_0^x
\frac{f(y)}{D_q}dy\right]
\end{equation}

where N is the normalization constant which can obtained as

\begin{equation}\label{2.22}
\int_{-\infty}^{+\infty}d\dot{x} \int_0^L dx \;P_{st}(x,\dot{x})=1
\end{equation}

It is easy to check that in the stationary state the mean velocity
is equal to zero;

\begin{equation}\label{2.23}
\langle\dot{x}\rangle_s=\int_{-\infty}^{+\infty}\dot{x}\;d\dot{x}\int_0^L
dx\; P_{st}(x,\dot{x})
\end{equation}

Several points are now in order: (i) Eq.(\ref{2.23}) suggests that
the stationary distribution (\ref{2.21}) is an equilibrium
distribution  because of the zero current condition. (ii) The
equilibrium distribution Eq.(\ref{2.21}) formally contains quantum
corrections to all orders in
$Q(x,\langle\delta\hat{x}^n\rangle)$. (iii) Since
$Q(x,\langle\delta\hat{x}^n\rangle)$ essentially arises due to
nonlinear part of the potential the nonlinearity and the quantum
effects are entangled in this quantity modifying the classical
part of the potential. Thus the classical force $-V'(x)$ is
modified by the quantum contribution. (iv) Since in the present
scheme one may express the quantum mechanical operator $\hat{x}=
x+\delta\hat{x}$ or $\dot{\hat{x}}=\dot{x}+\delta\dot{\hat{x}}$
where $x$ and $\dot{x}$ are quantum mechanical mean values and
$\langle\delta\hat{x}\rangle=\langle\delta\dot{\hat{x}}\rangle=0$
by construction and $[\delta \hat{x},\delta\dot{\hat{x}}]=i\hbar$
as noted earlier it follows that

\begin{equation}\label{2.24}
\langle\dot{\hat{x}}\rangle_{qs}=\langle\dot{x}+\delta\dot{\hat{x}}\rangle_{qs}=
\langle\dot{x}\rangle_s+\langle\langle\delta\dot{\hat{x}}\rangle\rangle_s=\langle\dot{x}\rangle_s
\end{equation}

The relation between three types of averages e.g.,
$\langle...\rangle_{qs}$, quantum statistical;
$\langle...\rangle_s$ statistical average over quantum mechanical
mean and $\langle...\rangle$, quantum mechanical mean must be
clearly distinguished. The relation Eq.(\ref{2.24}) expresses the
usual quantum current as a simple statistical average of the
quantum mechanical mean value in the present c-number scheme and
the decisive advantage of using this formalism is quite apparent.
(v) In absence of quantum correction term
$Q(x,\langle\delta\hat{x}^n\rangle)$ and $D_q\rightarrow
\frac{\gamma kT}{\overline{\Delta V}}$ as one approaches the
classical limit $(kT\gg \hbar\omega)$, the quantum Langevin
equation (\ref{2.18}) reduces to classical Langevin equation.
(vi) The zero current situation or equivalently the equilibrium
distribution function (\ref{2.21}) ensures the condition of
detailed balance in absence of any external driving. This
condition is a necessity in the present context and the formalism
since it guarantees that the quantum correction term does not
give any tilt or bring any asymmetry on the classical periodic
potential generating any unphysical current. This conclusion is
also true for an overdamped situation which we deal in the next
subsection.

\subsection{Equilibrium under overdamped condition}
Under overdamped condition the inertial term may be neglected and
one obtains from Eq. (\ref{2.18})

\begin{equation}\label{2.25}
\dot{x}=f(x)+\Gamma(t)
\end{equation}

where over-dot (.) refers to differentiation with respect to
dimensionless time $t$ defined as $ t=\frac{\overline{t}}{\tau_0}$
and $x=\frac{\overline{x}}{L}$. Therefore Eq.(\ref{2.25}) gives
the relation $\langle
\frac{d\overline{x}}{d\overline{t}}\rangle_s=\frac{L}{\tau_0}\langle\dot{x}\rangle_s=v_0\langle
f(x)\rangle_s$ or

\begin{equation}\label{2.26}
\langle\frac{d\overline{x}}{d\overline{t}}\rangle_s=v_0\int_0^1
f(x)P_{st}(x)dx
\end{equation}

Here we have denoted the characteristic velocity $v_0=L/\tau_0$.
The equation for probability density function $P(x,t)$
corresponding to Eq.(\ref{2.25}) is given by

\begin{equation}\label{2.27}
\frac{\partial P(x,t)}{\partial t}=-\frac{\partial
J(x,t)}{\partial x}
\end{equation}

where the probability current

\begin{equation}\label{2.28}
J(x,t)=f(x)P(x,t)-D_q\frac{\partial P(x,t)}{\partial x}
\end{equation}

In the stationary state $P(x)=Lt_{t\rightarrow\infty}P(x,t)$ , J
is constant as

\begin{equation}\label{2.29}
J=f(x)P(x)-D_q\frac{\partial P(x)}{\partial x}
\end{equation}

The solution of above equation for $P(x)$ reads as

\begin{equation}\label{2.30}
P(x)=-\frac{J}{D_q}\exp[-\psi(x)]\int_0^x\exp[\psi(y)]\;dy
+N\exp[-\psi(x)]
\end{equation}

where

\begin{equation}\nonumber
\psi(x)=-\int_0^x\frac{f(y)}{D_q}dy
\end{equation}

or $\psi(x)=\frac{V(x)-V(x)+\langle V(\hat{x})\rangle}{D_q}$ and
$N$ is constant. The last relation follows from (\ref{2.6}) and
(\ref{2.19}). Since $V(x)$ is periodic, \textit{i.e.}
$V(x)=V(x+1)$ we must have

\begin{equation}\label{2.31}
\psi(x)=\psi(x+1)
\end{equation}

For periodic boundary condition on (\ref{2.30}) and from
(\ref{2.31}) it follows that,

\begin{equation}\label{2.32}
\frac{J}{D_q}\int_x^{x+1}\exp[\psi(y)]\;dy =0
\end{equation}

Since the above integral is non-zero an overdamped Langevin
equation with periodic boundary condition shows $J=0$. This
corresponds to an equilibrium situation with probability density
function from (\ref{2.30}),

\begin{equation}\label{2.33}
P(x)=N\exp[-\psi(x)]
\end{equation}

Normalization constant N is $[\int_{0}^{1}\exp[-\psi(x)]]^{-1}$.
Therefore the quantum correction
$Q(x,\langle\delta\hat{x}^n\rangle)$ in $\psi(x)$, as expected,
can not break the detailed balance in the quantum system, nor the
symmetry of the potential. This conclusion is an important check
of the present formalism for a correct description of the
equilibrium.

\section{External noise-induced quantum transport}

Since at equilibrium detailed balance in the quantum stochastic
system under overdamped condition forbids any transport we
introduce an external noise on the system. The dynamics of the
particle is described by the equation

\begin{equation}\label{3.1}
\dot{x}=f(x)+\Gamma(t)+\xi(t)
\end{equation}

where $\Gamma(t)$ is the quantum internal noise of the bath with
the properties as noted earlier, $\xi(t)$ is a random telegraph
noise also known as dichotomous noise, which takes two possible
values $\xi(t)=\{-a,b\}$. If the probability of jumps per unit
time from one state are given by $P(-a\rightarrow b)=\mu_a$ and
$P(b\rightarrow-a)=\mu_b$ and if we assume $a\mu_b=b\mu_a$, then
this external stochastic process can be described by the first
two moments as

\begin{equation}\label{3.2}
\langle\xi(t)\rangle=0
\end{equation}

\begin{equation}\label{3.3}
\langle\xi(t)\xi(s)\rangle=\frac{Q_I}{\tau}\exp\left[-\frac{\mid{t-s}\mid}{\tau}\right]
\end{equation}

where the correlation time of the noise
$\tau=\frac{1}{\mu_a+\mu_b}$ and the noise intensity $Q_I=\tau a
b$, $\tau_a$ and $\tau_b$ are mean waiting times in the states $a$
and $b$ $(\mu_a=\frac{1}{\tau_a},\mu_b=\frac{1}{\tau_b})$
respectively. Therefore the three parameters intensity $Q_I$,
correlation time $\tau$, and asymmetry $\theta=b-a$ are the
characteristics of the noise. For symmetrical noise $a=b$. The
quantum nature of the problem therefore manifests itself in two
ways; first, through quantum corrections in $f(x)$ which we
consider in principle to all orders and secondly in quantum
diffusion coefficient $D_q$ for the noise of the bath. The
quantum equation of motion for joint probability densities can be
mapped into a classical problem by defining
\begin{equation}\nonumber
P_+(x,t)=P(x,b,t)\;\;\; ; \;\;\;P_-(x,t)=P(x,-a,t)
\end{equation}

so that Fokker-Planck equation with jump processes are given by

\begin{equation}\label{3.4}
\frac{\partial P_+(x,t)}{\partial t}=-\frac{\partial }{\partial
x}[f(x)+b]P_+(x,t)+D_q\frac{{\partial}^2}{{\partial
x}^2}P_+(x,t)-\mu_bP_+(x,t)+\mu_aP_-(x,t)
\end{equation}

\begin{equation}\label{3.5}
\frac{\partial P_-(x,t)}{\delta t}=-\frac{\partial }{\partial
x}[f(x)-a]P_-(x,t)+D_q\frac{{\partial}^2}{{\partial
x}^2}P_-(x,t)+\mu_bP_+(x,t)-\mu_aP_-(x,t)
\end{equation}

The total probability density $P(x,t)$ at any time is given by

\begin{equation}\label{3.6}
P(x,t)=P_+(x,t)+P_-(x,t)
\end{equation}

Eqs.(\ref{3.4}) and (\ref{3.5}) yield the equation of the motion
for $P(x,t)$ as

\begin{equation}\label{3.7}
\frac{\partial P(x,t)}{\partial t}=-\frac{\partial }{\partial
x}[f(x)]P(x,t)-\frac{\partial}{\partial
x}W(x,t)+D_q\frac{{\partial}^2}{{\partial x}^2}P(x,t)
\end{equation}

where $W(x,t)$ is an auxiliary distribution function

\begin{equation}\label{3.8}
W(x,t)=bP_+(x,t)-aP_-(x,t)
\end{equation}

which follows the equation

\begin{equation}\label{3.9}
\frac{\partial W(x,t)}{\partial t}=-\frac{\partial }{\partial
x}[f(x)+\theta]W(x,t)+D_q\frac{{\partial}^2}{{\partial
x}^2}W(x,t)-\frac{1}{\tau}W(x,t)-ab\frac{\partial}{\partial
x}P(x,t)
\end{equation}

The normalization conditions are

\begin{equation}\label{3.10}
\int_c^{c+1}P(x,t)dx=1\;\;\; ;\;\;\;   \int_c^{c+1}W(x,t)dx=0
\end{equation}

In the stationary state we obtain an expression for the constant
current $J$ as

\begin{equation}\label{3.11}
-D_qP'(x)+f(x)P(x)+W(x)=J
\end{equation}
and also we have
\begin{equation}\label{3.12}
\tau D_qW''(x)-\tau[\theta+f(x)]W'(x)-[1+\tau f'(x)]W(x)=Q_IP'(x)
\end{equation}

$P(x)$ and $W(x)$ are the stationary solutions of the coupled
equations (\ref{3.7}) and (\ref{3.9}). It is difficult to solve
analytically the above two  equations for arbitrary potential. In
what follows we consider the solutions under two specific cases
$(a)$ large correlation time and $(b)$ small correlation time.

\subsection{Large correlation time}

We return to Eqs.(\ref{3.11}) and (\ref{3.12}) and rewrite
(\ref{3.12}) as

\begin{equation}\label{3.13}
D_q\frac{d^2W(x)}{dx^2}-\frac{d}{dx}\{\theta+f(x)\}W(x)-\frac{1}{\tau}W(x)
=ab\frac{dP(x)}{dx}
\end{equation}

For $\tau\gg 1$ we neglect the term with $\frac{1}{\tau}$.
Integration over Eq. (\ref{3.13}) then leads to

\begin{equation}\label{3.14}
D_q W'(x)-[\theta+f(x)]W(x)=abP(x)+D
\end{equation}

$D$ is constant. We now put the equilibrium solution for $P(x)$,
 Eq.(\ref{2.33}) in (\ref{3.14}) and solve it for $W(x)$ to obtain
(we put $D=0$ to make the system free from bias due to external
fluctuating force averaged over a period)

\begin{equation}\label{3.15}
W(x)=\exp[-\psi_1(x)]\left\{\frac{abN}{D_q}\int_0^x
\exp{}\psi_2(y)dy +C_m \right\}
\end{equation}

\begin{equation}\label{3.16}
C_m =-\frac{abN}{D_q}\;\frac{\int_{c_0} ^{c_0+1}
\exp[-\psi_1(x)]\int_0^x \exp[\psi_2(y)]\;dy\; dx}{\int_{c_0}
^{c_0+1} \exp[-\psi_1(x)]\;dx}
\end{equation}

\begin{equation}\label{3.17}
\psi_1(x)=-\int_0^x \frac{\theta+f(y)}{D_q}\;dy
\end{equation}

\begin{equation}\label{3.18}
\psi_2(x)=-\int_0^x \frac{\theta}{D_q}\;dy
\end{equation}

\begin{equation}\label{3.19}
\psi(x)=-\int_0^x \frac{f(y)}{D_q}\;dy
\end{equation}

Here $N$ is given by normalization constant in (\ref{2.33}).
Putting the solutions for $W(x)$ and $P(x)$ (\ref{3.15} and
\ref{2.33}) in (\ref{3.11}) as a first approximation we obtain the
expression for current as the lowest order iterative solution
which is given by

\begin{equation}\label{3.20}
J=\frac{\int_x^{x+1}\exp[-\psi_2(y)]\;\{\frac{abN}{D_q}
\int_0^y\exp[\psi_2(z)]\;dz +C_m
\}\;dy}{\int_x^{x+1}\exp[-\psi(y)]\;dy}
\end{equation}

The expression for current is valid for large correlation time of
the dichotomous  noise but formally takes into consideration of
quantum effects to all orders. In order to check the consistency
of the above expression we now examine the following limiting
situations. First, we consider the dichotomous noise to be
symmetric, $\textit{i.e.}$, $\theta=0$. $J$ then reduces to

\begin{equation}\label{3.21}
J=\frac{\int_x^{x+1}\{\frac{abN}{D_q}y+C_m
\}\;dy}{\int_x^{x+1}\exp[-\psi(y)]\;dy}
\end{equation}

and
\begin{equation}\label{3.22}
C_m =-\frac{abN}{D_q}\;\frac{\int_{c_0} ^{c_0+1} x
\exp[-\psi(x)]\;dx}{\int_{c_0} ^{c_0+1} \exp[-\psi(x)]\;dx}
\end{equation}

If now $V(x)$ is assumed to be of inversion symmetric, then
$\langle V(\hat{x})\rangle$ is also symmetric and $C_m$ would be
zero and $J=0$ in such situation since

\begin{equation}\label{3.23}
\int_{-1/2} ^{+1/2} x \exp[-\psi(x)]\;dx=0
\end{equation}

\begin{equation}\label{3.24}
\int_{-1/2} ^{+1/2} \frac{abN}{D_q}\;x\; dx=0
\end{equation}

Therefore with symmetric potential and symmetric dichotomous
noise, the current is zero even in the presence of quantum
corrections. To obtain a quantum current it is necessary that
either the periodic potential should be asymmetric and/or noise
$\xi(t)$ should be asymmetric and  vice-versa.

 We now proceed to analyze the current
under non-equilibrium condition and the related quantum effects.
One of the prime quantities for this analysis is the potential
$V(x)$ or the corresponding force term $f(x)$ given by

\begin{equation}\label{3.25}
f(x)=-[V'(x)- Q(x
,\langle\delta\hat{x}^n\rangle)]=-\frac{\partial}{\partial
x}[V(x)+\sum_{n\geq 2}\frac{1}{n!}V^n(x) \langle
\delta\hat{x}^n\rangle]
\end{equation}

The quantum correction terms can be determined as follows. We
return to the operator equation (\ref{2.2}) and put
$\hat{x}(\overline t)=\overline{x}(\overline t)+
\delta\hat{x}(\overline t)$ and $\hat{p}(\overline
t)=\overline{p}(\overline t)+ \delta\hat{p}(\overline t)$ where
$\overline{x}(\overline t)=\langle\hat{x}(\overline{t})\rangle$
and $\overline{p}(\overline
t)=\langle\hat{p}(\overline{t})\rangle$ are the quantum mechanical
mean values of the operators $\hat{x}$ and $\hat{p}$
respectively. By construction
$[\delta\hat{x},\delta\hat{p}]=i\hbar$ and $\langle\delta
\hat{x}\rangle=\langle\delta \hat{p}\rangle=0$. We then obtain the
quantum correction equation

\begin{equation}\label{3.26}
m \delta\ddot{\hat{x}}+\int^{\overline{t}}_0
d\overline{t}'\gamma(\overline{t}-\overline{t}')\delta\dot{\hat{x}}(\overline{t'})
+\overline{V}''(\overline{x})\delta\hat{x}+ \sum_{n\geq
2}\frac{1}{n!}\overline{V}^{n+1}(\overline x)(\delta\hat{x}^n-
\overline{\langle \delta\hat{x}^n\rangle})=
\hat{\Gamma}(\overline{t})-\overline{\Gamma}(\overline{t})
\end{equation}

Again in the overdamped limit we discard  the inertial term
$m\delta \ddot{\hat{x}}$. We then perform a quantum mechanical
average with initial product separable coherent states of the
oscillators of the bath only to get rid of the internal noise term
and to obtain the reduced operator equation for the system as

\begin{equation}\label{3.27}
\gamma\delta\dot{\hat{x}} +\overline{V}''(\overline{
x})\delta\hat{x}+ \sum_{n\geq
2}\frac{1}{n!}\overline{V}^{n+1}(\overline x)(\delta\hat{x}^n-
\langle \overline{\delta\hat{x}^n\rangle})=0
\end{equation}

With the help of (\ref{3.27}) we then obtain the equations for
$\overline{\langle\delta\hat{x}^n(t)\rangle}$

\begin{equation}\label{3.28}
\frac{d}{d\overline
t}\overline{\langle\delta\hat{x}^2\rangle}=\frac{1}{\gamma}\left[-2\overline
V''(\overline x) \overline{\langle\delta\hat{x}^2\rangle}
-\overline V'''(\overline
x)\overline{\langle\delta\hat{x}^3\rangle}\right]
\end{equation}

\begin{equation}\label{3.29}
\frac{d}{d\overline
t}\overline{\langle\delta\hat{x}^3\rangle}=\frac{1}{\gamma}\left[-3\overline
V''(\overline x)\overline{\langle\delta\hat{x}^3\rangle}
-\frac{3}{2}\overline V'''(\overline
x)\overline{\langle\delta\hat{x}^4\rangle}+\frac{3}{2}\overline
V'''(\overline x)\overline{\langle\delta\hat{x}^2\rangle}^2\right]
\end{equation}

and so on. Taking into account of the lowest order contribution
$\overline{\langle\delta\hat{x}^2\rangle}$ explicitly we may write

\begin{equation}\label{3.30}
d\overline{\langle\delta\hat{x}^2\rangle}=-\frac{2}{\gamma}\overline
V''(\overline x)\overline{\langle\delta\hat{x}^2\rangle}
d\overline t
\end{equation}

The overdamped deterministic motion gives $\gamma d\overline
x=-\overline V'(\overline x)dt$ which when used in (\ref{3.30})
yields after integration

\begin{equation}\label{3.31}
\overline{\langle\delta\hat{x}^2\rangle}=\Delta_q[\overline
V'(\overline x)]^2
\end{equation}

where
$\Delta_q=\frac{\overline{\langle\delta\hat{x}^2\rangle}_{x_c}}{[\overline
V(\overline x_c)]^2}$ and $\overline x_c$ refers to a known
reference point. Eq.(\ref{3.31}) results in

\begin{equation}\label{3.32}
\overline f(\overline x)=-[\overline V'(\overline x)+\Delta_q
\overline V'''(\overline x)[\overline V'(\overline x)]^2]
\end{equation}

We now emphasize an important point. If the potential is
symmetric, then the quantum correction in Eq.(\ref{3.32}) is an
odd function just as $V'(x)$. This implies that quantum correction
to classical potential has not destroyed the inversion symmetry of
$V(x)$. Thus the approximation in deriving the leading order
quantum effect is consistent with symmetry requirement of the
problem.

To illustrate the nature of current we now consider a symmetric
cosine potential with period $2\pi$

\begin{equation}\label{3.33}
\overline V(\overline x)=\frac{1}{2}(\cos\overline x+1)
\end{equation}

The force terms and other related quantities $\psi_1(x)$,
$\psi_2(x)$ and $\psi(x)$ after scaling, take the following forms:

\begin{equation}\label{3.34}
f(x)=\pi[\sin2\pi x-\Delta_{q}\sin^32\pi x]
\end{equation}

\begin{equation}\label{3.35}
\psi_1(x)=-[\Delta_1\cos^32\pi x+\Delta_2\cos2\pi
x+\frac{\theta}{D_q}x+\Delta_3]
\end{equation}

\begin{equation}\label{3.36}
\psi_2(x)=-\frac{\theta}{D_q}x
\end{equation}

\begin{equation}\label{3.37}
\psi(x)=-[\Delta_1\cos^32\pi x+\Delta_2\cos2\pi x+\Delta_3]
\end{equation}

\begin{equation}\label{3.38}
\Delta_1=-\frac{\Delta_q}{6D_q};\;\;\;\;\;\Delta_2=\frac{\Delta_q-1}{2D_q}
;\;\;\;\;\;\Delta_3=\frac{1 -\frac{2}{3}\Delta_q }{2D_q}
\end{equation}

In Fig.(1) we illustrate the variation of current for a fixed
value of the system non-linearity $\Delta_q(=0.04)$ and asymmetry
parameter $\theta(=1.0)$ as a function of quantum diffusion
coefficient $D_q$ . One observes that with increase of $D_q$ the
magnitude of current increases to a maximum followed by a
decrease. For a fixed $D_q$ with increase of the strength of the
external dichotomous noise (proportional to the product $ab$) the
current increases.The effect of quantization of a classical
ratchet is shown in Fig.(2), where we make a comparison of the
current vs temperature profile for the classical and the quantum
$(\Delta_q=0.3)$ cases for $a=1.75$, $ b=2.75$. One observes that
in the low temperature region the classical current is
significantly lower in magnitude than the quantum current, and at
high temperature the effect of quantization become insignificant.

\subsection{Short correlation time}

In the regime of short correlation time $\tau \ll 1$ of
dichotomous noise we follow Kula \textit{et al} \cite{kul} to
expand $P(x)$, $W(x)$ and $J$ in power series with $\tau$ as a
smallness parameter;

\begin{equation}\label{3.39}
P(x)=\sum_{n=0}^{\infty} \tau^n P_n(x)\;\;\;;\;\;\;
W(x)=\sum_{n=0}^{\infty} \tau^n W_n(x)\;\;\;and
\;\;\;J=\sum_{n=0}^{\infty} \tau^n J_n
\end{equation}

Making use of the above expressions in (\ref{3.11}) and
(\ref{3.12}) we obtain the following set of equations,

\begin{equation}\label{3.40}
-\;D_q P'(x) +f(x)P_n(x)+W_n(x)=J_n
\end{equation}

\begin{equation}\label{3.41}
W_n(x)=D_q W''_{n-1}(x)-[\theta+f(x)]
W'_{n-1}(x)-f'(x)W_{n-1}(x)-Q_I P'_n(x)
\end{equation}

with $n=1,2,3...$

\begin{equation}\nonumber
W_0(x)=-Q_I P'_0(x)
\end{equation}

The probability functions $P_n(x)$ obey the periodicity
conditions and they are normalized over dimensionless period
$(L=1)$. We thus obtain the zero order contributions as

\begin{eqnarray}
J_0 & = & 0\label{3.42}\\
P_0(x) &=& N \exp{\left[ \int_0^x
\frac{f(y)}{D_q+Q_I}\;dy\right]}\label{3.43}
\end{eqnarray}

with normalization constant

\begin{equation}\label{3.44}
N^{-1}=\int_0^1\exp{\left[ \int_0^x
\frac{f(y)}{D_q+Q_I}\;dy\right]} dx
\end{equation}

The higher order contributions can be obtained following Kula
\textit{et al}. For the present purpose the leading order current
is given by

\begin{eqnarray}
J_1 &=& \left[\int_0^1 N^{-1} P_0 (x) dx \int_0^1 N P^{-1}_0(x) dx
\right]^{-1}\nonumber\\
&\times& \left[ \frac{\theta Q_I}{(D_q+Q_I)^2} \int_0^1 f^2(x) dx
+\frac{Q_I^2}{(D_q+Q_I)^3}\int_0^1 f^3(x) dx \right]\label{3.45}
\end{eqnarray}

The key quantity for the above equation is the force term $f(x)$
with leading order nonlinear correction (\ref{3.32}). For the
symmetric smooth cosine potential of the form
$V(x)=\frac{1}{2}[\cos 2\pi x +1]$ as given by (\ref{3.33}),
$f(x)$ is an odd function [Eq.\ref{3.34}]. For asymmetric
dichotomous fluctuations $(\theta \neq 0)$ and symmetric
potential the leading order current is proportional to the
integral $f^2(x)$; the integral over $f^3(x)$ being zero. On the
other hand the current is proportional to the integral over
$f^3(x)$ for symmetric$(\theta=0)$ dichotomous noise. Therefore
we have shown that in the short correlation time limit it is not
possible to obtain any noise induced transport with symmetric
noise and symmetric potential.

For $f(x)$ given by (\ref{3.34}) one obtains explicitly the
quantum current

\begin{equation}\label{3.46}
J_1=\frac{\pi^2}{2 I_1\;I_2}\;\frac{\theta Q_I}{(D_q+Q_I)^2}
\left[ \frac{5}{8} \Delta_q^2-\frac{3}{2} \Delta_q+1 \right]
\end{equation}

where

\begin{eqnarray}
I_1 &=& \int_0^1 \exp\left[ -\int_0^x \frac{f(y)}{(D_q+Q_I)}\; dy
\right] dx\label{3.47}\\
\nonumber\\
I_2 &=& \int_0^1 \exp\left[ \int_0^x \frac{f(y)}{(D_q+Q_I)}\; dy
\right] dx\label{3.48}\\
\nonumber\\
\int_0
^x \frac{f(y)}{(D_q+Q_I)} &=& \Delta_4 \cos^3 2\pi x +
\Delta_5 \cos 2\pi x + \Delta_6\label{3.49}
\nonumber\\
\Delta_4 &=& -\frac{\Delta_q}{6
(D_q+Q_I)}\;\;\;;\;\;\;\Delta_5=\frac{(\Delta_q-1)}{2(D_q+Q_I)}\label{3.50}\\
\nonumber\\
\Delta_6 &=& \frac{(1-\frac{2}{3}\Delta_q)}{2
(D_q+Q_I)}\label{3.51}
\end{eqnarray}

We now numerically illustrate the behaviour of quantum current
given by Eq.(\ref{3.46}).  The effect of quantization of the
reservoir is apparent in Fig.(3) in the variation of current with
$Q_I$ for several values of quantum diffusion coefficient $D_q$ of
the heat bath for fixed $\Delta_q(=0.04)$, $\theta(=1.0)$ and
$\tau(=0.1)$. For small $D_q$, the current falls off monotonically
after reaching maxima. The maxima and the current drops for higher
values of quantum diffusion coefficient since thermalization
prevails over the dynamics, in general. In Fig.(4) we compare the
current vs temperature profile for the classical and the quantum
$(\Delta_q=0.3)$ cases for fixed $a(=1.0)$, $ b(2.0)$ and
$\tau(=0.1)$. We observe again that in the low temperature range
the current is significantly higher for the quantum case.

In order to examine the influence of the correlation time $\tau$
of the nonthermal noise on current we plot in Fig.(5) the
variation of current as a function of $D_q$ for several values of
$\tau $ for fixed values of $a(=1.0)$, $b(=2.0)$, $\theta(=1.0)$
and $\Delta_q(=0.04)$. All the bell-shaped curves exhibit maxima
at optimal $D_q$ values. Increase in correlation time $\tau$
results in enhancement of directed motion, and shift of the
maxima, towards the origin. Physically this implies that departure
from equilibrium is increasingly favored for larger correlation
time of the external noise in this region.

 The form for noise induced-current in the limit of short correlation time
 of the stochastic dichotomous process expresses the functional
 dependence of the current on $\tau$, $f$,
$D_q$ and $Q_I$, through Eq.(\ref{3.46}) as $J=\tau J_1$. It is
interesting to comment on the effect of noise statistics of the
stochastic processes on the current. This aspect has been
explained carefully by Doering \textit{et al}\cite{doe} sometime
back in a classical context by considering a family of stochastic
processes known as "Kangaroo processes"  parameterized  by an
another probability distribution $p(z)$. Based on singular
perturbation expansion in $\tau$ it has been shown that the
expression for the current is modified by a prefactor which
depends on the moments of the distribution $p(z)\;\;(\langle z^n
\rangle = \int z^n p(z)dz)$. It is apparent that consideration of
such processes in the present quantum mechanical context is
expected to give rise such a prefactor since the expression for
current in the short correlation time limit in classical and
quantum mechanical cases has a common structure.

\section{conclusion}

 Based on the traditional system-reservoir model we have formulated
 the quantum stochastic dynamics in the overdamped limit and
 analyzed the problem of a ratchet device with an external,
 exponentially correlated asymmetric dichotomous noise. The governing
 equations of motion are classical looking in form but quantum
 mechanical in their content. The quantization of the dynamics is
 manifested in two different ways. First, the harmonic oscillator
 reservoir is quantum mechanical in character and its internal
 noise characteristics and the fluctuation-dissipation relation are
 described by the canonical thermal Wigner distribution.
  The special advantage of using this
 c-number distribution (where width $D_q$ corresponds to the
 strength of fluctuation of the thermal bath) is that
 it remains a valid positive definite pure state distribution even
 at absolute zero. This allows the theory to be extended to the deep
 tunneling region which remains normally inaccessible in many
 theoretical treatments. Secondly, the nonlinearity of the potential
  brings in additional quantum contribution since the nonlinear
   terms of the potential beyond the harmonic one are entangled with
   quantum corrections. Therefore the system experiences an
   effective force term $f(x)$ comprising a classical $-V'(x)$
   plus a quantum correction term $Q(x ,\langle\delta\hat{x}^n\rangle)$
   as $f(x)=-V'(x)+Q(x ,\langle\delta\hat{x}^n\rangle)$ this
     consideration leads us to the form of generalized equilibrium
   distribution in terms of a nonlocal potential $\psi(x)$ as $P \sim
   \exp[\psi(x)]$ where $\psi(x)=\int_0^x \frac{f(x')}{D_q}dx'$.
   The implication of this factor in ratchet effect or in Landauer
   blow-torch effect has been thoroughly examined \cite{mag,kam,doe,kul} by
   Van Kampen, B\"{u}ttiker and others. A close look into the
   expression for current in Eq.(\ref{3.20}) or Eq.(\ref{3.46})
   reveals that the origin of fluctuation induced current
   essentially rests on this factor and therefore the contribution
   of the nonlinearity induced quantum effect on this current
   becomes quite apparent. It is important to note that the
   essential
   requirements demanded by symmetry considerations and
   thermodynamic consistency condition have been fulfilled in our
   treatment. We now summarize the main conclusions of this study:

   (i) We note that quantization can not break the symmetry of the
   ratchet device but, in general, may change the superposition of
   amplitudes of the periodic nonlinear function so that the current
   is significantly affected. This is apparent from the structure
   of the force term $f(x)$ pointed out earlier. For example, for a pendulum potential
   $V(x)=\frac{1}{2}(\cos{x}+1)$, which has been used for superionic conductors,
   $f(x)$ [$=-V'(x)-\Delta_q V'''(x)\left(V'(x)\right)^2$ for the leading
   order quantum correction where $\Delta_q\sim O(h^2)$] is a superposition
   of $\sin{x}$ and $\Delta_q \sin^3{x}$, a classical and a quantum
   part, respectively. It therefore follows from the previous discussions
   that the quantum part of potential affects $\psi(x)$ and consequently
   the current.

   (ii) Our formulation offers simple analytical solutions in two
   limiting cases, large and small correlation times of the
   telegraphic noise. We observe that while at low temperature
   quantization significantly enhances the classical current, at higher
   temperature the difference is insignificant. This may be
   interpreted in terms of an interplay between the quantum
   diffusion coefficient $D_q$ and the force term $f(x)$ appearing
   in the effective potential $\psi(x)$ as $\int_0^x
   \frac{f(x')}{D_q}dx'$. As the temperature $T \rightarrow 0$,
   $D_q$ approaches to the value $
    \frac{1}{2}\hbar\omega_0$, the vacuum limit and also in the
     deep tunneling region the anharmonic terms in $f(x)$ do not
     contribute significantly, the integrand
    increases sharply. On the other hand, as the temperature
    increases, $D_q$ increases resulting in a decrease of
    $\psi(x)$. Since for the classical current, the quantum
    contribution to $f(x)$ is absent, one observes a crossover of
    the  of the classical and the quantum current in an intermediate region of the
    temperature, where $D_q$ and $f(x)$ compete with each other and
    beyond which the quantum current is marginally lower than the
    classical current. For a further increase of temperature the
    classical and the quantum current merge identically, as
    expected.

    (iii) We have examined the role of the strength of the external
    noise and the correlation time in the generation of noise-induced
     current. These factors, as usual are instrumental in moderating
     the statistical stationary state of the nonequilibrium system
     that lacks detailed balance. A possible experimental
     realization of the
     distinctive behaviour of a quantum ratchet in contrast to its
     classical counterpart in a superionic conductor driven by a
     dichotomous noisy electric field at low temperature has been suggested.

     Our study is confined to analytical solutions in the two limits
     of correlation time of the external noise.
      A direct numerical simulation of quantum
     ratchet using c-number quantum Langevin equation with
     dichotomous noise over the entire range of correlation time
     is necessary to compliment these observations. Furthermore
     since the problem of quantum Brownian dynamics in a periodic
     (pendulum like) potential covers a wide area comprising
     Josephson super-current through a tunneling junction,
     Bloch-wall motion, Frochlich superconductors etc., we
     expect the formulation to be useful in a wider theoretical
     context. Application of the theory in related issues are also
     worth-pursuing and will be addressed elsewhere.

\acknowledgments Thanks are due to the CSIR, Govt. of India, for a
fellowship (PKG) and for partial financial support (grant No:
01/(1740)/02/EMR II).

\newpage
\begin{center}
{\bf Figure Captions}
\end{center}

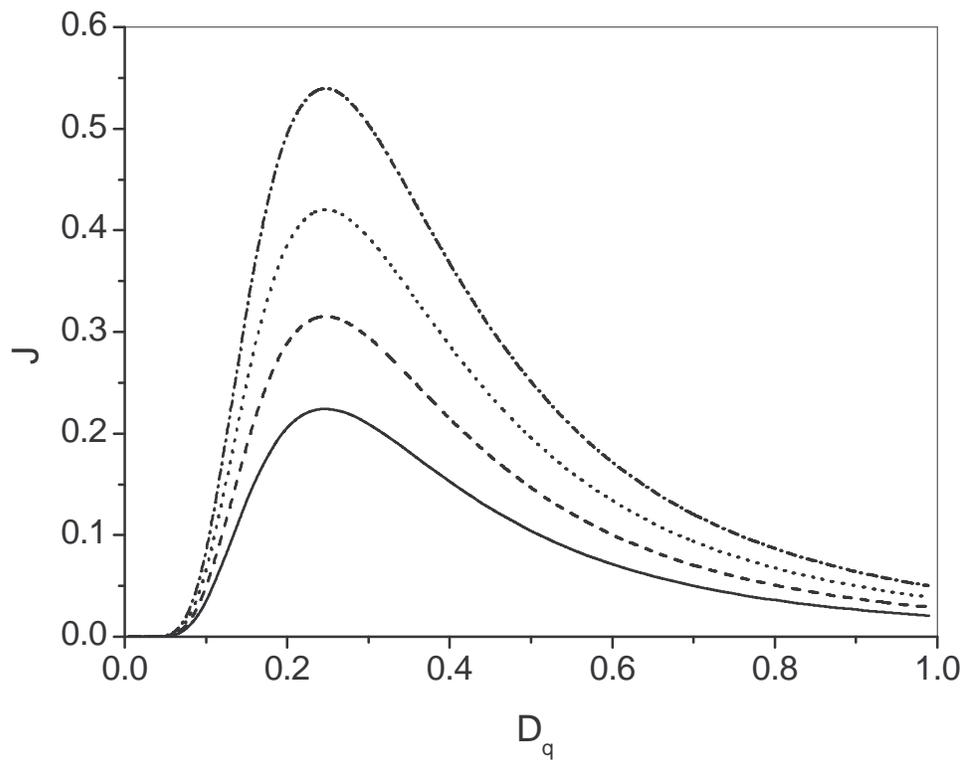
Fig.1: Variation of current ($J$) with quantum diffusion
coefficient $D_q$ in the large correlation time limit for
$\Delta_q=0.04$, $\theta=1.0$ and (i) $a=1.0$ and $b=2.0$ (solid
line), (ii) $a=1.25$ and $b=2.25$ (dashed line), (iii) $a=1.5$ and
$b=2.5$ (dot line) and (iv) $a=1.75$ and $b=2.75$ (dash-dot line).

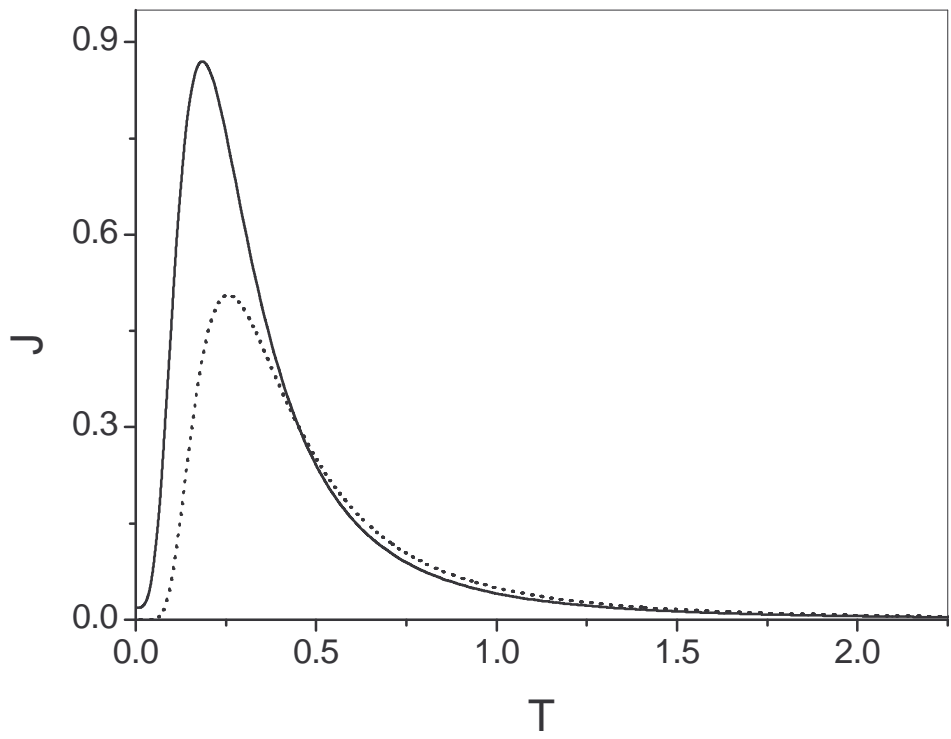
Fig.2: Comparison of quantum and classical current ($J$) vs
temperature ($T$) profile for the parameter set $a=1.75$,
$b=2.75$, $\Delta_q=0.0$ (classical, dotted line) and
$\Delta_q=0.3$ (quantum, solid line) in the large correlation time
limit.

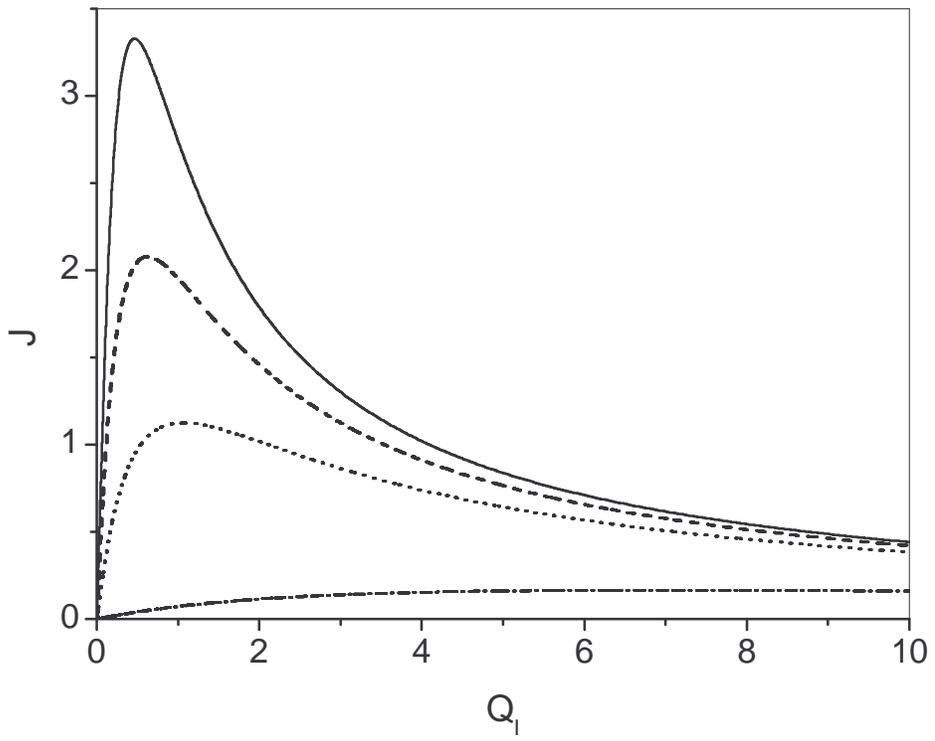
Fig.3: Variation of current ($J$) with nonthermal noise strength
($Q_I$) for different values of quantum diffusion coefficients
$D_q=0.25$ (solid line), $D_q=1.5$ (dashed line), $D_q=3.0$
(dotted line) and $D_q=7.0$ (dash-dot line) for the parameter set
$\Delta_q=0.04$, $\theta=1.0$ and $\tau=0.01$ in the short
correlation time limit.

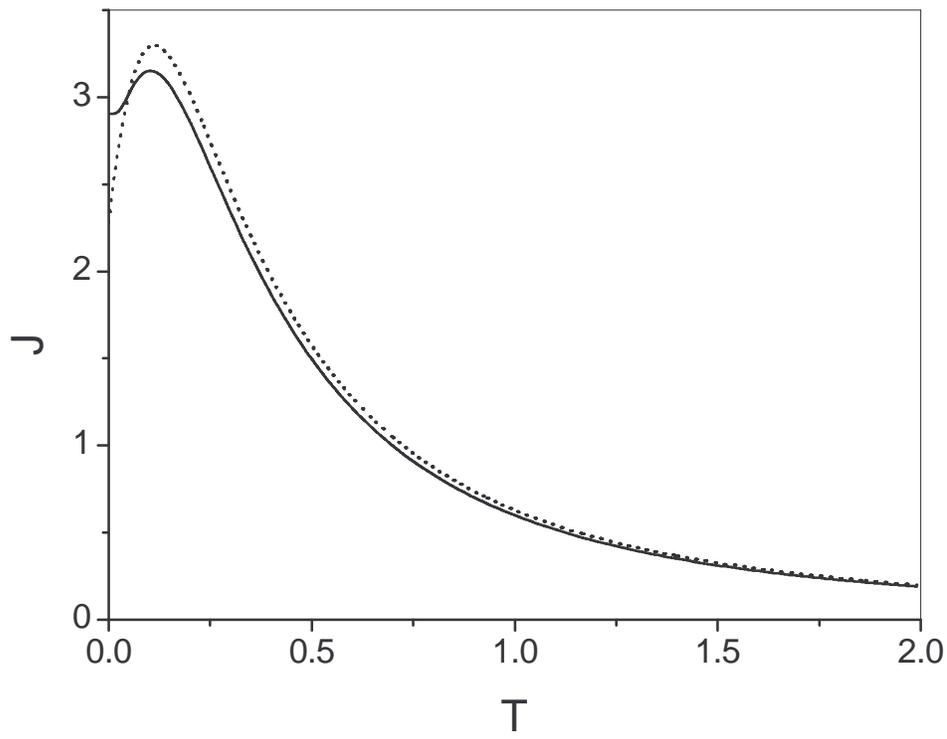
Fig.4: Comparison of quantum and classical current ($J$) vs
temperature ($T$) profile for the parameter set $a=1.0, b=2.0$,
$\tau=0.1$ and $\Delta_q=0.0$ (classical, dotted line) and
$\Delta_q=0.3$ (quantum, solid line) in the short correlation time
limit.

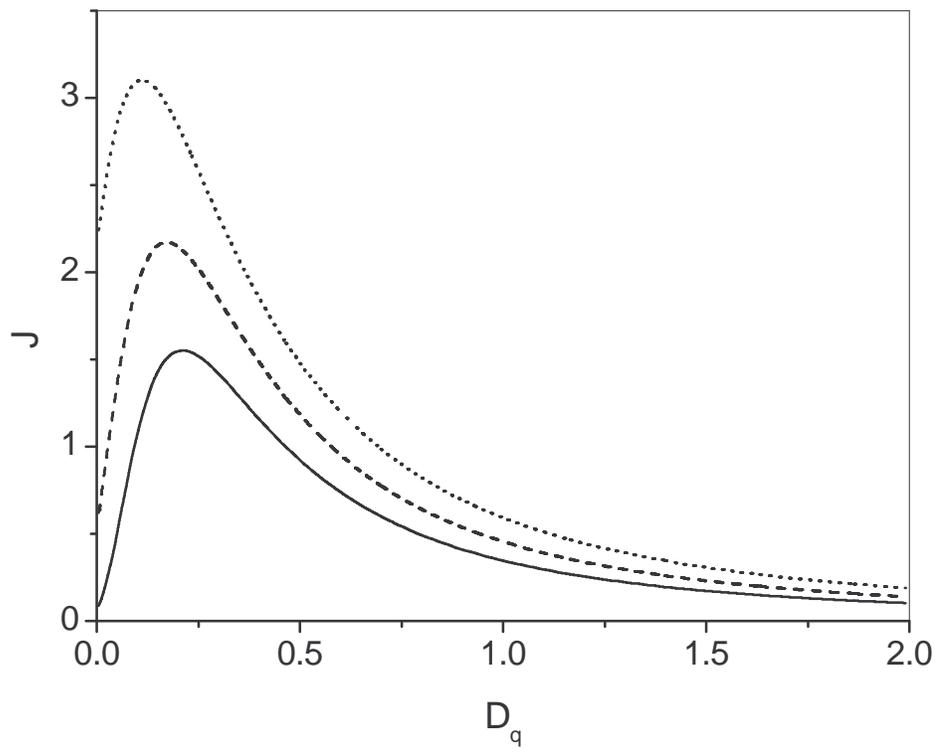
Fig.5: Variation of current ($J$) with quantum diffusion
coefficient for different value of correlation time $\tau$ of
nonthermal noise $\tau=0.05$ (solid line), $\tau=0.07$ (dashed
line) and $\tau=0.1$ (dotted line) for the parameter set
$\Delta_q=0.04$, a=1.0, b=2.0 and $\theta=1.0$ in the short
correlation time limit.
\end{document}